\documentclass[aps,prl,twocolumn,notitlepage,nofootinbib]{revtex4-1}
\usepackage{amsmath,amsfonts,amssymb,bm,cancel}
\usepackage{graphicx,float}
\usepackage{enumitem}
\usepackage{color}
\usepackage{soul}

\begin{document}

\title{Metastable state en route to traveling-wave synchronization state}

\author{Jinha Park}
\affiliation{CCSS, CTP and Department of Physics and Astronomy, Seoul National University, Seoul 08826, Korea}
\author{B. Kahng}
\email{bkahng@snu.ac.kr}
\affiliation{CCSS, CTP and Department of Physics and Astronomy, Seoul National University, Seoul 08826, Korea}

\begin{abstract}
The Kuramoto model with mixed signs of couplings is known to produce a traveling-wave synchronized state. Here, we consider an abrupt synchronization transition from the incoherent state to the traveling-wave state through a long-lasting metastable state with large fluctuations. Our explanation of the metastability is that the dynamic flow remains within a limited region of phase space and circulates through a few active states bounded by saddle and stable fixed points. This complex flow generates a long-lasting critical behavior, a signature of a hybrid phase transition. We show that the long-lasting period can be controlled by varying the density of inhibitory/excitatory interactions. We discuss a potential application of this transition behavior to the recovery process of human consciousness.
\end{abstract}

\maketitle

A hybrid phase transition (HPT) is a discontinuous transition that accompanies critical phenomena. Recent hybrid percolation model studies~\cite{multi,zhou,kcore_mendes,kcore,golden} have discovered that the system stays at a long-lasting metastable preparatory step on the way to an explosive transition, during which the so-called powder keg is accumulated~\cite{explosive}. In this regard, one may wonder if there exists a similar metastable state in a synchronization transition. However, the presence of a metastable state has been rarely highlighted in synchronization problems~\cite{kurths}. In this paper, we reveal that such an intermediate metastable state indeed exists on the way to a discontinuous synchronization transition near the hybrid critical point. Moreover, we show that this long-lasting metastable step can be understood as persisting circulation inside a metastable basin, characterized by balancing between saddle points and stable fixed points. 

The Kuramoto model~\cite{kuramoto,book1,book2,book3,Winfree80} has been successfully used to investigate the properties of the synchronization transition (ST) and is expressed as 
\begin{align}
\dot\theta_i = \omega_i + \frac{K}{N}\sum_{k=1}^{N}\sin (\theta_k-\theta_i),
\end{align}
where $\theta_i$ denotes the phase of each oscillator $i$; $\omega_i$ is the intrinsic frequency of an oscillator $i$, which follows a distribution $g(\omega)$; $K$ is the coupling constant; and $N$ is the number of oscillators in the system. STs are characterized by a complex order parameter defined as $Z(t)\equiv \sum_{i=1}^{N}e^{i\theta_i}/N = Re^{i\Psi(t)}$, where $R$ is the magnitude of the phase coherence; $R=0$ for the incoherent (IC) state, and $R\neq 0$ for the coherent (C) state. For a usual Gaussian $g(\omega)$, a continuous ST occurs at the critical coupling strength $K_c$. We instead use a uniform $g(\omega)$ that exhibits an abrupt ST~\cite{Winfree80} with a post-jump criticality $\beta=2/3$~\cite{Pazo05}. We remark that the bimodal $g(\omega)$ gives a first-order transition; however, it is not hybrid~\cite{Martens09}.

Here, the Kuramoto model with uniform $g(\omega)$ is extended to a mixture of two opposite-sign coupling constants $K_1<0$ and $K_2>0$ to the fraction $1-p$ and $p$, motivated by excitatory and inhibitory couplings in neural networks~\cite{kopell,Hong11}. This extension further distinguishes the C phase into $\pi$ and traveling wave (TW) phases, and is characterized by the steady rotation of the complex angle of the order parameter $\Omega\sim\Psi/t$; $\Omega=0$ in the $\pi$ state, whereas $\Omega \ne 0$ in the TW phase. Hereafter, we call our model the competing Winfree--Paz\'{o} (c-WP) model~\cite{Pazo05,Winfree80,kopell,Hong11}, where $\omega$ and $K$ of an oscillator follow the probability distribution
\begin{eqnarray}
g(\omega,K)=\frac{1}{2\gamma}\Theta(\gamma-|\omega|)\big[(1-p)\delta(K-K_1)+p\delta(K-K_2)\big],\nonumber \\
\end{eqnarray}
where $\Theta$ represents the Heaviside step function. The coherent steady state is characterized by two groups of oscillators separated by an angle $\Delta$ in the phase space $\theta$ that correspond to the inhibitory and excitatory populations. When $\Delta=\pi$ ($\pi$ state), the two groups are balanced and the steady-state rotation $\Omega=\Psi/t$ is zero. When $\Delta\neq\pi$, the TW order with $\Omega\neq 0$ emerges. 

\begin{figure}
\centering
\includegraphics[width=1\linewidth]{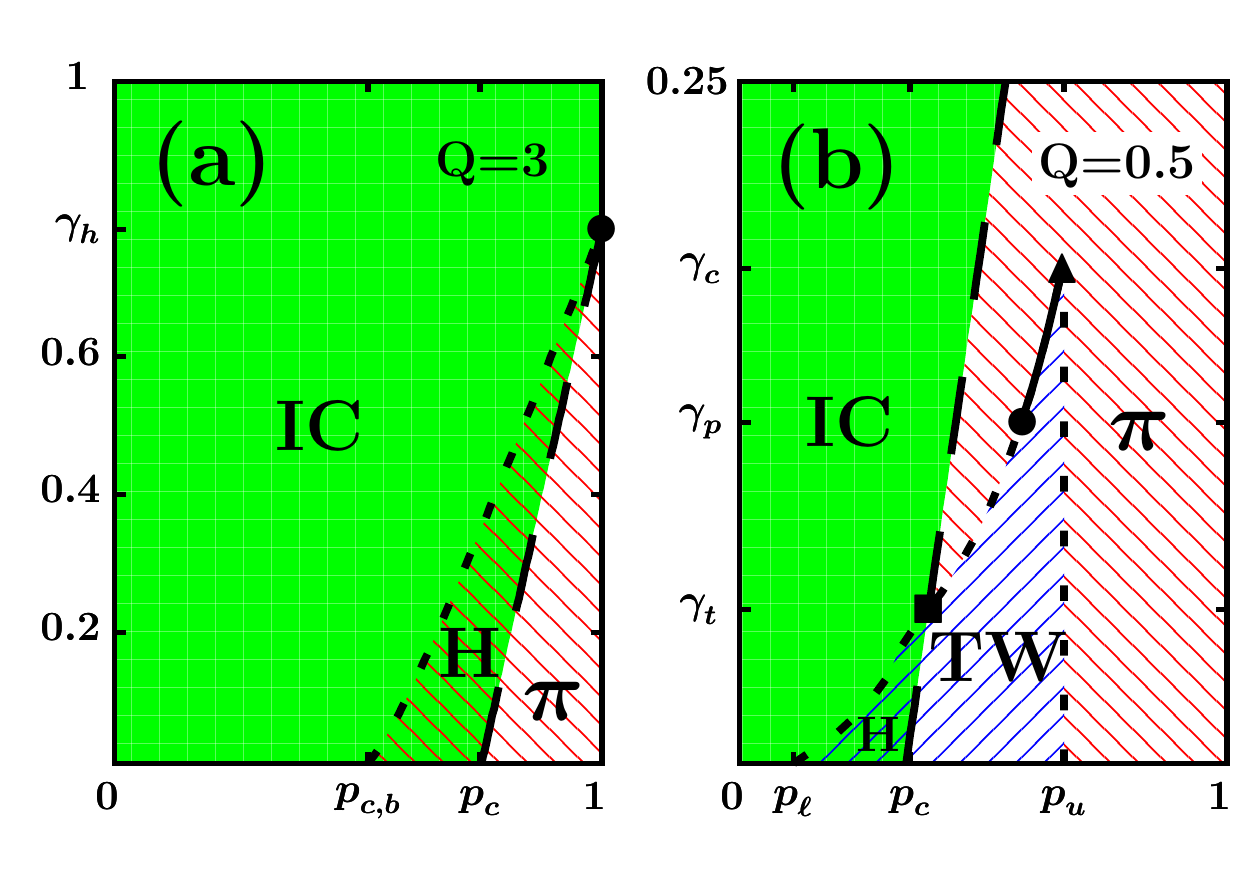}
\caption{(Color online) Phase diagram of synchronization transitions (STs) in the $(p,\gamma)$ plane. $p$ is the fraction of oscillators with $K_2 > 0$ and $\gamma$ is the half width of the uniform distribution $g(\omega)$. The phase diagram contains the IC and $\pi$ phases in (a) and (b), and the TW phase when in (b). The solid line represents a second-order transition, and both types of dashed lines represent first-order transitions, but the transition from the IC phase to the $\pi$ phase is hybrid. {\bf H} represents a hysteresis zone. The symbol $\bullet$ at $\gamma_h\approx 0.78$ in (a) corresponds to the hybrid critical point of the Winfree--Paz\'{o} (WP) model. The symbols $\blacktriangle$ at $\gamma_c\approx 0.18$, $\bullet$ at $\gamma_p\approx 0.13$, and $\blacksquare$ at $\gamma_t\approx 0.064$ in (b) represent critical points across which different types of phases or phase transitions emerge. The TW phase is absent when $Q>1$, or $Q<1$ and $\gamma> \gamma_c$.}
\label{fig:phaseDiagram}
\end{figure}

\begin{figure*}
\centering
\includegraphics[width=1\linewidth]{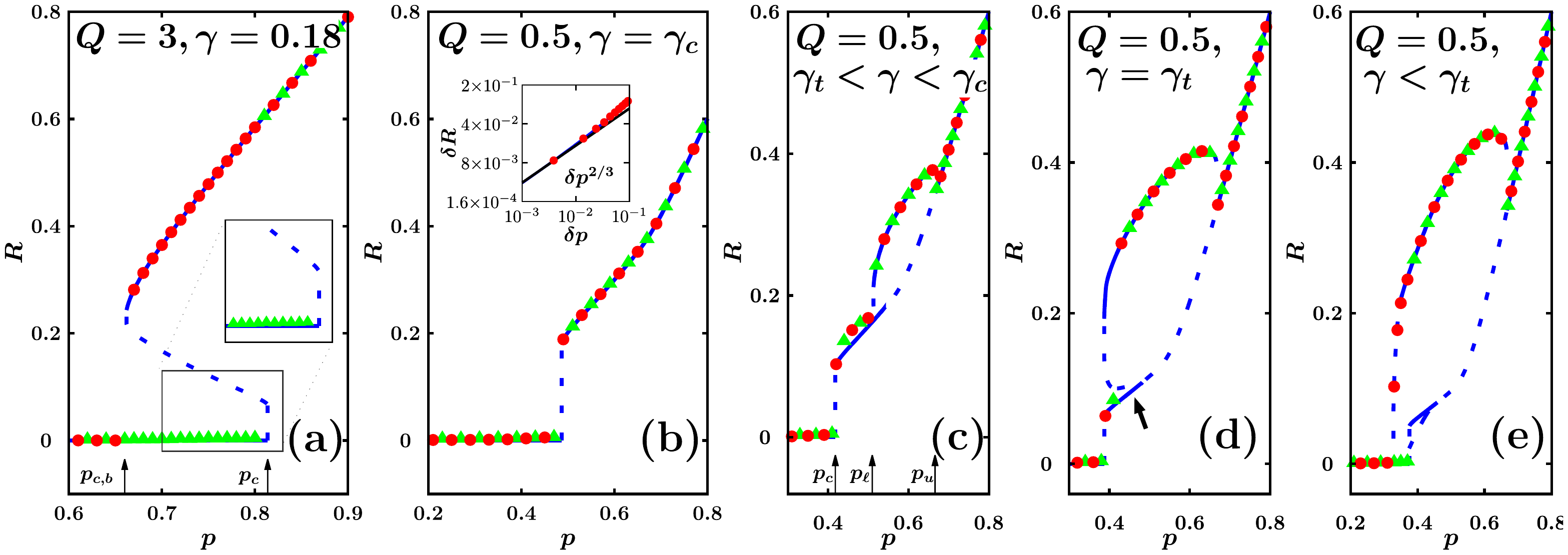}
\caption{ (Color online) Diverse types of STs at $Q=3$ and $Q=0.5$. Green triangles and red circles denote data points of $R(p)$ obtained from simulations starting from the IC and C initial states, respectively. Solid(dashed) blue curves are self-consistency solutions representing stable(unstable) states, according to the stability criterion, {Eq.\eqref{eq:stability}}. In (a), a first-order transition and hysteresis occur between $p_c$ and $p_{c,b}$. The inset shows a close-up near $p_c$. We emphasize that the unstable line does not continuously drop to $p_c$, but instead has a finite gap of size $\gamma/|K_1|$. In (b), a hybrid phase transition (HPT) occurs with the critical exponent $\beta_p=2/3$ at $p_c$. A close check of the exponent value is shown in the inset. The black line guides a slope of $2/3$. (c) The TW phase emerges at $\gamma_c$ and exists in the range $[p_\ell,p_u]$. When $\gamma_t<\gamma<\gamma_c$, ${\rm IC}\leadsto\pi\rightarrow{\rm TW}\rightarrow\pi$ occur with increasing $p$. (d) At $\gamma=\gamma_t,p_c=p_\ell$; thus, ${\rm IC}\dashrightarrow{\rm TW}\rightarrow\pi$ occur. The part of the $\pi$ line (indicated by arrow) that is stable according to the criterion is actually metastable. (e) When $\gamma<\gamma_t$ $(\gamma=0.05)$, $p_\ell<p_c<p_u$. $R$ jumps from the IC state to the TW state, and a hysteresis occurs between the IC and TW states at $[p_\ell,p_c]$, where ${\rm IC}\dashrightarrow{\rm TW}\rightarrow\pi$ occurs. Different types of arrows distinguish the types of phase transitions: continuous ($\rightarrow$), discontinuous ($\dashrightarrow$), and hybrid ($\leadsto$).}
\label{fig:op}
\end{figure*}

We construct a self-consistency equation of the c-WP model to obtain the steady-state order parameter solutions $(R,\Omega)$ and perform numerical simulations to verify their stabilities. Unexpectedly, a rich phase diagram involving the HPT is obtained, as shown in Fig.~\ref{fig:phaseDiagram}. When $Q<1$, a supercritical HPT occurs and the behavior of the order parameter is expressed as 
\begin{equation}
R(p)=\left\{
\begin{array}{lr}
0 & ~{\rm for}~~  p < p_c, \\[2pt]
R_c+a(p-p_c)^{\beta_p} & ~{\rm for}~~ p\geq p_c,
\end{array}
\right.
\label{eq:order_1}
\end{equation}
near the hybrid critical point $R_c=\gamma/K_2$ and $p_c= [Q+4\gamma/(\pi K_2)]/(Q+1)$, with a noninteger exponent $\beta_p=2/3$. When $Q>1$, the critical exponent $\beta_p$ remains the same while the post-jump branch has the opposite direction and becomes unstable (Fig~\ref{fig:op}(a)). The transition from the IC phase to the $\pi$ phase is first-order and exhibits a hysteresis curve in the region between $p_c$ and $p_{c,b}$. Notice that this subcritical HPT is different from the usual subcritical Hopf bifurcation. The unstable line in the inset of Fig.~\ref{fig:op}(a) does not continuously drop to $p_c$, but instead has a finite gap of size $\gamma/|K_1|$. On the other hand, when $g(\omega)$ is Lorentzian~\cite{Hong11}, either a continuous transition or a discontinuous transition occurs, and without any critical behaviors or presence of the metastable states. 

\begin{figure*}
\centering
\includegraphics[width=1\linewidth]{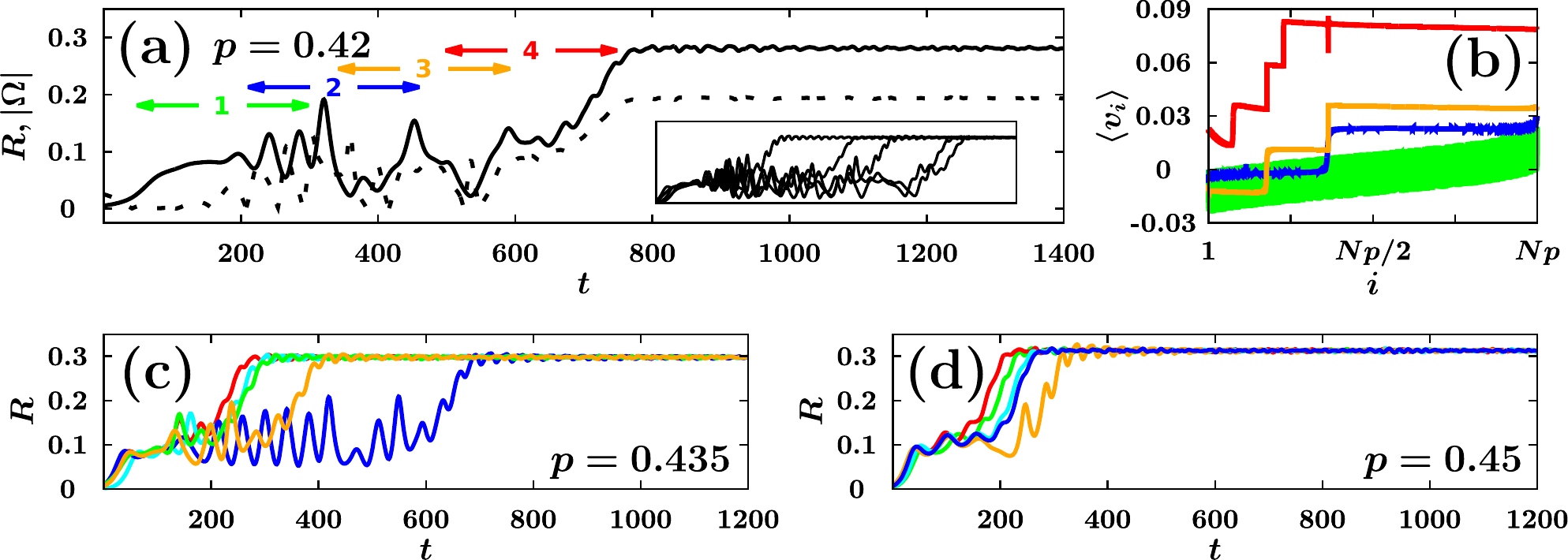}
\caption{ (Color online) Tiered ST from the IC state to TW state through the metastable $\pi$ state. $R(t)$ was obtained at various $p$ for the system size $N=25\,600$, $Q=0.5$, and $\gamma=0.064$. (a) At $p=0.42$, the TW state appears as the steady state, and the state does as metastable. The solid and dashed lines correspond to $R(t)$ and $|\Omega(t)|$, respectively. Note that both the temporal and sample-to-sample (inset) fluctuations of $R$ are large during the metastable period. In (b), the velocities of $K_2$ oscillators are averaged over each specified time interval, as indicated by the corresponding colors and cluster numbers in (a). The oscillators are indexed in ascending order of the intrinsic frequencies. We find several intermediate states with different numbers of clusters composed of oscillators with similar velocities. The number of clusters increases as the stages proceed. In (a), (c), and (d), as $p$ is increased, the metastable period becomes shorter. Subsequently, the TW state is reached shortly.}
\label{fig:metastable}
\end{figure*}

It is intriguing to check the stability of the self-consistency solution. To perform this task, the so-called empirical stability criterion proposed in Ref.~\cite{stefanovska} was checked on the c-WP model. The stability matrix $\hat{S}$ of Ref.~\cite{stefanovska} is reproduced as follows:
\begin{align}
\begin{pmatrix} \dot{\delta R} \\ \dot{\delta \psi} \end{pmatrix}
&= A \begin{pmatrix}(\partial_R F_R)-1 & R^2 \partial_\Omega F_R \\ R^{-1}\partial_R F_\Omega & R\partial_\Omega F_\Omega\end{pmatrix} \begin{pmatrix} \delta R \\ \delta \psi \end{pmatrix} \nonumber \\
&\equiv A \hat{S}\begin{pmatrix} \delta R \\ \delta \psi \end{pmatrix}\\
F_R(R,\Omega) &\equiv \int_{\rm locked} dKd\omega g(\omega,K) 
\sqrt{1-\left(\omega/KR\right)^2} \nonumber \\
F_\Omega(R,\Omega) &\equiv \int_{\rm drifting} dKd\omega g(\omega,K) 
\sqrt{\left(\omega/KR\right)^2-1}
\label{eq:stability}
\end{align}
where $F_R$ and $F_\Omega$ correspond to the real and imaginary parts of the self-consistent order parameter. The system is (empirically) stable if and only if $\textrm{tr}(\hat{S})<0$ and $\det(\hat{S})>0$.
The result is presented by the blue solid (stable) and dashed (unstable) curves in Figs.~\ref{fig:op}. Our numerical result suggests that this linear stability criterion is partly fulfilled; some portions of the ``stable'' $\pi$ curve are not covered by the simulation data points in the long-time limit. Interestingly, the order parameter stays for quite a long time at these uncovered parts, before it finally settles down in the stable stationary line occupied by the symbols in Figs.~\ref{fig:op}(d)--\ref{fig:op}(e). These parts uncovered by simulation data are not stable but metastable. Fig.~\ref{fig:metastable}(a) shows the dynamic phase transition just above the hybrid critical point $p_c$; a tiered ST occurs from the IC phase to the TW phase through a long-lasting metastable $\pi$ phase. The order parameter $R$ exhibits large temporal and sample-to-sample fluctuations in this metastable interval. As $p$ is increased further, the fluctuations decrease and the metastable period becomes shorter (Figs.~\ref{fig:metastable}(c) and (d)). Subsequently, the metastability is lost and the ST to the TW state occurs directly. These behaviors terminate at $p_u$.

\begin{figure*}
\centering
\includegraphics[width=1\linewidth]{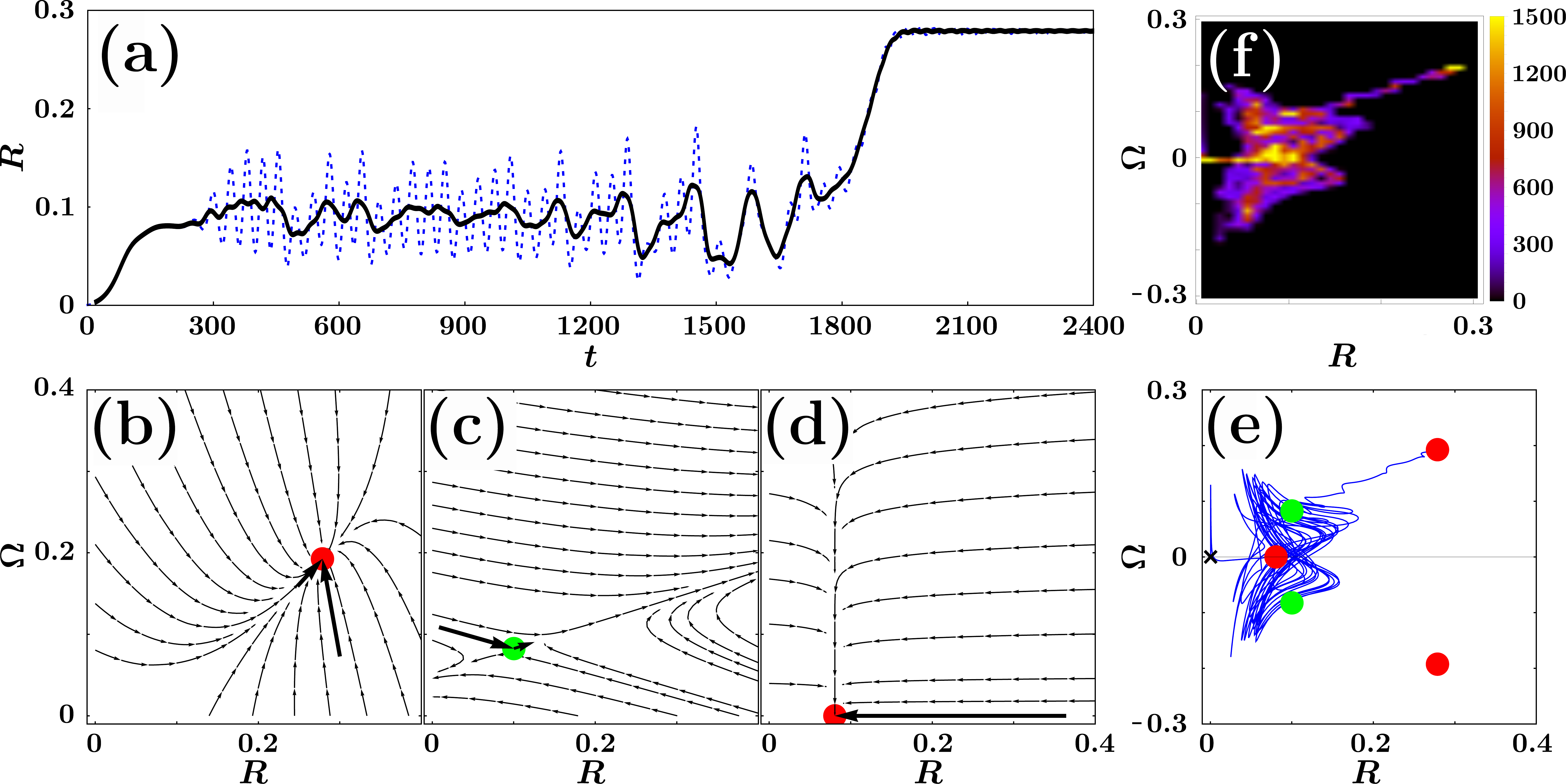}
\caption{ (Color online) The flow of the order parameter in the two-step synchronization transition. (a) Plot of blue dotted curve $R(t)$ vs $t$ at $p=0.418$. The time-averaged black curve $\langle R \rangle$ is obtained using a sliding $40$\,-s time window centered at each $t$ with a window step of $1$\,s. (b)--(d) The linearized flow in the $(R,\Omega)$ plane. Two stable points of $\pi$ and TW states are represented by red circles, and a saddle point of the TW state is shown in green. (e) An actual flow is obtained from simulations. $\bm{x}$ represents the starting point. (f) Frequencies of the dynamic flow passing through each state in the phase space. A few states (yellow) are active throughout the flow.}
\label{fig:twostep}
\end{figure*}

The empirical linear flows given by Eq.~\eqref{eq:stability} around each of the steady-state solutions $(R,\Omega)$ are shown in Figs.~\ref{fig:twostep}(b)--\ref{fig:twostep}(d). We remark that all TW solutions in Figs.~\ref{fig:twostep}(b) and \ref{fig:twostep}(c) exist in pairs owing to the symmetry $\Omega\leftrightarrow-\Omega$. The red circle in Fig.~\ref{fig:twostep}(b) represents a TW stable point, the green circle in Fig.~\ref{fig:twostep}(c) represents a TW saddle point, and the red circle in Fig.~\ref{fig:twostep}(d) represents a $\pi$ state with neutral stability. In Fig.~\ref{fig:twostep}(d), the eigenvalue in the $\Omega$ direction is extremely small compared with that of the $R$ direction. Thus, the corresponding eigenvector in the vertical $\Omega$ direction can be effectively understood as a $\dot\Omega\approx 0$ nullcline. The dotted blue line in Fig.~\ref{fig:twostep}(a) and the blue line in Fig.~\ref{fig:twostep}(e) correspond to a trajectory $(R(t),\Omega(t))$ realized from simulation. In Fig.~\ref{fig:twostep}(e), the system passes by the $\pi$ state of Fig.~\ref{fig:twostep}(d) and is then attracted by the saddle point of Fig.~\ref{fig:twostep}(c), forming unstable oscillations. It stays for a long time in the metastable basin bounded by the $\Omega$ nullclines and the saddle point. After escaping from the region, the dynamics flows immediately into the stable TW point. We remark that this trajectory is in fact a two-dimensional projection of a higher-dimensional dynamics and all other degrees of freedom do not vanish, inducing dynamic noise, until the stable steady TW state is finally reached posterior to the escapement.

Numerical simulations are performed using the fourth-order Runge--Kutta method with $\Delta t=0.01$. The number of oscillators is $N=25\,600$ and total runtimes are over $t=10^4$\,s, sufficiently longer than the transient periods. Fluctuations in $R$ and $\Omega$ at the stationary state were averaged out over the last 10\% of total runtime. The stationary state may additionally depend on the initial coherence, especially in the hysteresis zone. Oscillator phases are randomly assigned either in the range $[0,2\pi]$ or $[0,\pi/100]$, corresponding to the initially coherent or incoherent state. Natural frequencies of oscillators with $K_1 < 0$ and $K_2 > 0$ are regularly sampled between $[-\gamma,\gamma]$. $K_2$ is set to unity for convenience, leading to $K_1=-Q$. 

The two-step jump transition of Figs.~\ref{fig:metastable} and \ref{fig:twostep}(a) near the hybrid critical point $(p_c,\gamma_t)$ closely resembles those observed in the percolation on interdependent networks~\cite{multi,zhou}, $k$-core percolation~\cite{kcore_mendes,kcore} and the two-step contagion model~\cite{golden} near the critical point of the HPT. In those systems, the order parameters also show a long-lasting plateau with large fluctuations, as we observed in the metastable states of the c-WP model. 
During this lengthy period, the system accumulates a so-called powder keg for the later explosive transition~\cite{explosive,golden}. This feature is also similar to the accumulation of similar-size clusters near the transition point of the restricted percolation model, which exhibits a HPT~\cite{r_percolation}. During the metastable period of the c-WP model, the excitatory $K_2$ oscillators form a number of velocity clusters, i.e., clusters with similar velocities, when averaged over short time intervals, as shown in Fig.~\ref{fig:metastable}(b).  The number of clusters discretely increases as the dynamics proceeds. However, the divisions into small $K_2$ clusters are transient. Eventually, those clusters merge into the largest cluster and become monolithic in the TW state, whereas the inhibitory $K_1$ clusters break off and become liquid. It would be interesting to find out whether those $K_2$ clusters play an equivalent role as a time bomb that sets off an abrupt escapement of the metastable basin.

The c-WP synchronization model may have potential applications to the recovery dynamics of human consciousness from anesthetic-induced unconsciousness~\cite{Alkire,Hudson}. Inhibitory anesthetics such as $\gamma$-aminobutyric acid hinder cortical synchronization and the brain in turn loses its ability to integrate information, vigilance, and responsiveness~\cite{Alkire}.
Recent electroencephalogram (EEG) experiments have revealed that the power spectrum of the cortical local field potentials during the conscious state peak at a certain intrinsic frequency~\cite{Hudson}. This feature may be interpreted as an indicator of the TW synchronization in the c-WP model. The consciousness recovery dynamics of the anesthetic-induced brain passes through a sequence of several discrete activity states. Moreover, transitions between those metastable states are abrupt~\cite{Hudson}. A series of studies have previously modeled the anesthetic recovery using Kuramoto-type synchronization models~\cite{Lee17,Lee18}. However, our model further involves the metastable dynamic restoration of coherence by the discrete merging of velocity clusters. More interestingly, it deals with excitatory and inhibitory neural interactions through a controllable parameter $p$ and the recovery period is reduced by increasing $p$ beyond a threshold $p_c$, corresponding to the clinical findings that the recovery time is reduced with lesser anesthetic concentration. We remark that reducing the inhibitory anesthetic concentration also corresponds to increasing $p$ of our model. Moreover, our analysis not only provides a visualization scheme but also opportunities to manipulate the metastable terrain directly by controlling the saddle-point position in the phase space. 

In summary, we found that near the critical point of the HPT, a tiered ST occurs from the IC state to the TW state through the intermediate $\pi$ state. The dynamic process in the metastable state was explained as the circulating flow through a few active states in the phase space, which exhibits large temporal and sample-to-sample fluctuations. We discussed that such a tiered ST can be a potential model for the process by which the brain recovers from pathological states to the awake state.

\begin{acknowledgments}
This work was supported by the National Research Foundation of Korea by Grant No. NRF-2014R1A3A2069005.
\end{acknowledgments}


\begin{thebibliography}{99}
\bibitem{multi} S. V. Buldyrev, R. Parshani, G. Paul, H. E. Stanley and S. Havlin, Nature (London) {\bf 464,} 1025 (2010).
\bibitem{zhou} D. Zhou, A. Bashan, R. Cohen, Y. Berezin, N. Shnerb and S. Havlin, Phys. Rev. E {\bf 90}, 012803 (2014).
\bibitem{kcore_mendes} G. J. Baxter, S. N. Dorogovtsev, K. E. Lee, J. F. F. Mendes and A. V. Goltsev, Phys. Rev. X {\bf 5,} 031017 (2015).
\bibitem{kcore} D. Lee, M. Jo, and B. Kahng, Phys. Rev. E {\bf 94,} 062307 (2016).
\bibitem{golden} D. Lee, W. Choi, J. Kert\'esz and B. Kahng, Sci. Rep. {\bf 7}, 5723 (2017).
\bibitem{explosive} R. M. D'Souza and J. Nagler, {\sl Nat. Phys.} {\bf 11}, 531 (2015).
\bibitem{kurths} P. Ji, T. K. DM. Peron, P. J. Menck, F. A. Rodrigues and J. Kurths, {\sl Phys. Rev. Lett.} {\bf 110}, 218701 (2013).
\bibitem{Winfree80} A. T. Winfree, {\sl The Geometry of Biological Time} (Springer, Berlin, 1980).
\bibitem{kuramoto} Y. Kuramoto, in {\sl International Symposium on Mathematical Problems in Theoretical Physics}, edited by H. Araki, Lecture Notes in Physics Vol. 30 (Springer, New York, 1975).
\bibitem{book1} S. H. Strogatz, {\sl Sync: The Emerging Science of Spontaneous Order} (Hyperion, New York, 2003).
\bibitem{book2} G. V. Osipov, J. Kurths and C. Zhou, {\sl Synchronization in Oscillatory Networks} (Springer, Berlin, 2007).
\bibitem{book3} S. Boccaletti, {\sl The Synchronized Dynamics of Complex Systems}, (Elsevier, Oxford, U.K., 2008).
\bibitem{Pazo05} D. Paz\'o, {\sl Phys. Rev. E} {\bf 72}, 046211 (2005).
\bibitem{Martens09} E. A. Martens, E. Barreto, S. H. Strogatz, E. Ott, P. So and T. M. Antonsen, {\sl Phys. Rev. E} {\bf 79}, 026204 (2009).
\bibitem{kopell} C. B\"orgers, and N. Kopell, {\sl Neural Comput.} {\bf 15}, 509 (2003).
\bibitem{Hong11} H. Hong and S. H. Strogatz, {\sl Phys. Rev. Lett.} {\bf 106}, 054102 (2011).
\bibitem{stefanovska} D. Iatsenko, S. Petkoski, P. V. E. McClintock and A. Stefanovska, {\sl Phys. Rev. Lett.} {\bf 110}, 064101 (2013).
\bibitem{r_percolation} Y. S. Cho, J. S. Lee, H. J. Herrmann and B. Kahng, {\sl Phys. Rev. Lett.} {\bf 116,} 025701 (2016). 
\bibitem{Alkire} M. T. Alkire, A. G. Hudetz, G. Tononi, {\sl Science}~ {\bf 322}, 876 (2008).
\bibitem{Hudson} A. E. Hudson, D. P. Calderon, D. W. Pfaff, A. Proekt, {\sl Proc. Natl. Acad. Sci. U.S.A.} {\bf 111}, 9283 (2014).
\bibitem{Lee17} J. Y. Moon, J. Kim, T. W. Koh, M. Kim, Y. Iturria-Medina, J. H. Choi, J. Lee, G. A. Mashour, and U. Lee, {\sl Sci. Rep.} {\bf 7}, 46606 (2017).
\bibitem{Lee18} U. Lee, M. Kim, K. Lee, C. M. Kaplan, D. J. Clauw, S. Kim, G. A. Mashour and R. E. Harris, {\sl Sci. Rep.} {\bf 8}, 243. (2018)
\end{thebibliography}
\end{document}